\pdfoutput=1
\documentclass[a4paper]{article}
\usepackage{graphicx}
\usepackage{mathptmx}      
\usepackage{amssymb}
\usepackage{graphicx}
\usepackage{multirow}	
\usepackage{multicol}
\usepackage{url}
\usepackage[margin=1.65in]{geometry}
\graphicspath{{./img/}}

\begin{document}

\title{Centrality in the Global Network of Corporate Control}


\author{Frank W. Takes         \and
        Eelke M. Heemskerk 
}

\title{Centrality in the Global Network \\ of Corporate Control}
\author{
Frank W. Takes and Eelke M. Heemskerk \\ \url{{takes,e.m.heemskerk}@uva.nl} \vspace{2mm} \\
CORPNET, University of Amsterdam \\ 
\url{http://corpnet.uva.nl} }
\date{May 22, 2016}

\maketitle


\begin{abstract}
Corporations across the world are highly interconnected in a large global network of corporate control. 
This paper investigates the \emph{global board interlock network}, covering $400,000$ firms linked through $1,700,000$ edges representing shared directors between these firms.   
The main focus is on the concept of \emph{centrality}, which is used to investigate the embeddedness of firms from a particular country within the global network.
The study results in three contributions.
First, 
to the best of our knowledge for the first time we can investigate the topology as well as the concept of centrality in corporate networks at a global scale, allowing for the largest cross-country comparison ever done in interlocking directorates literature. 
We demonstrate, amongst other things, extremely similar network topologies, yet large differences between countries when it comes to the relation between economic prominence indicators and firm centrality. 
Second, 
we introduce two new metrics that are specifically suitable for comparing the centrality ranking of a partition to that of the full network.  
Using the notion of \emph{centrality persistence} we propose to measure the persistence of a partition's centrality ranking in the full network. 
In the board interlock network, it allows us to assess the extent to which the footprint of a national network is still present within the global network. 
Next, the measure of \emph{centrality ranking dominance} tells us whether a partition (country) is more dominant at the top or the bottom of the centrality ranking of the full (global) network. 
Finally, comparing these two new measures of \emph{persistence} and \emph{dominance} between different countries allows us to classify these countries based the their embeddedness, measured using the relation between the centrality of a country's firms on the national and the global scale of the board interlock network. \\

\end{abstract}

\newpage


\section{Introduction}
\label{sec:introduction}

\begin{figure}[t]   
    \centering
    \includegraphics[width=\textwidth]{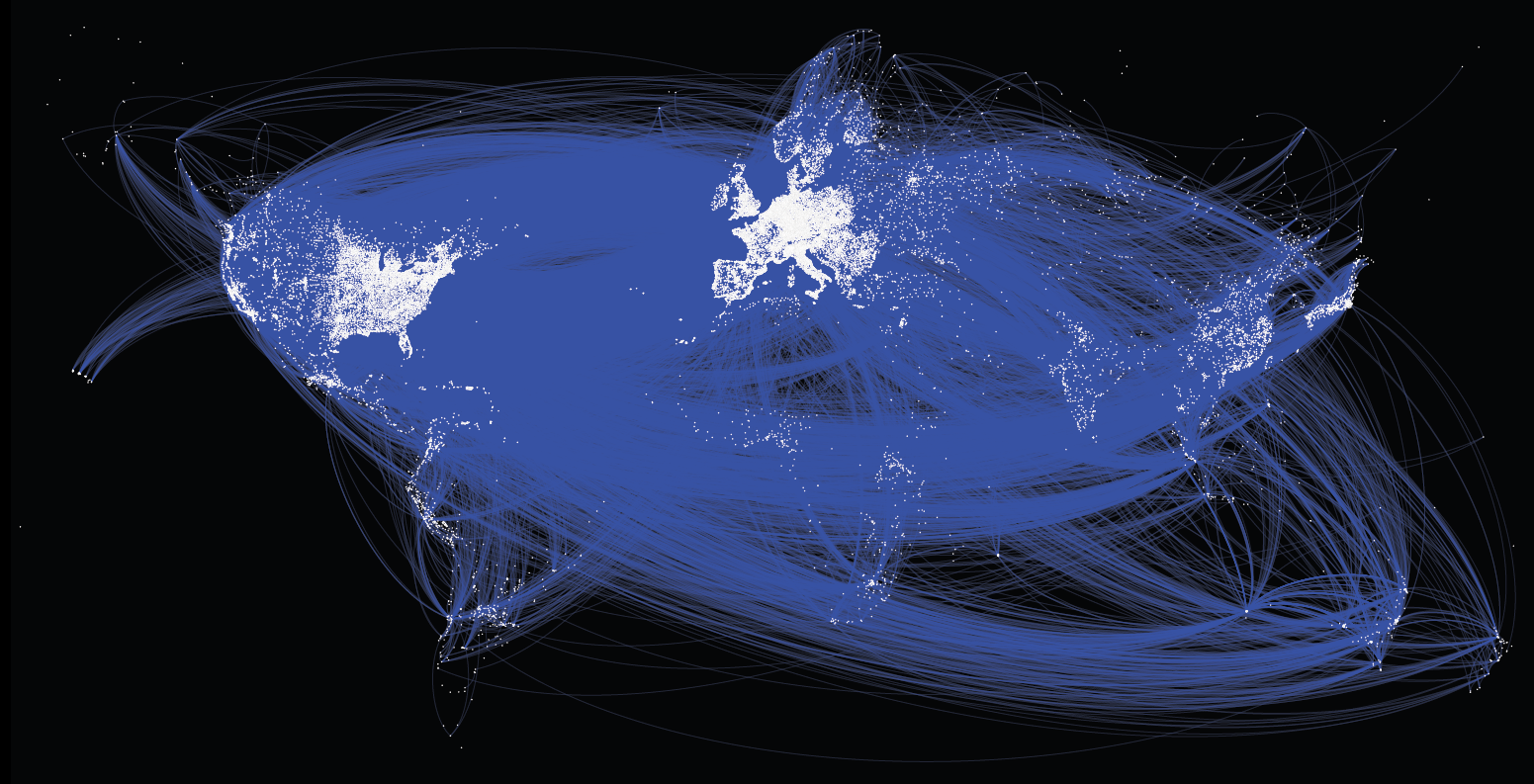}  
    \caption{Geographical visualization of the global board interlock network, consisting of around $400,000$ firms and over $1,700,000$ board interlocks. Visualized using Gephi (\url{http://gephi.org}).}   
    \label{fig:connectedfirms}
\end{figure}

Although often depicted as atomistic and individualistic market actors, corporations are tightly embedded in networks of power and control. 
Foundational elements of these networks are interlocking directorates, where officers of one firm also serve on the board of another firm. 
Increasingly, these hitherto national business communities' networks now form a global network of corporate control~\cite{heemskerk2015,heemskerktakes2015,heemskerktakes2016soc,kogut2012small,vitali2011network}. 
We refer to this structure of interlocking directorates as the global \emph{board interlock network}, which is in fact an undirected network (graph) consisting of firms (nodes) and particular relationships (edges) between these firms. 
Two firms are connected if they share a common senior level director, officer or board member, essentially modeling the social ties that exist between firms. 
Our global corporate network is based on nearly $400,000$ firms and more than $1,700,000$ board interlock ties between firms. 
A geographical visualization of this network is given in Figure~\ref{fig:connectedfirms}, illustrating the dense interconnectedness of our global economy through board interlocks.

A network modeling approach allows the use of existing metrics and techniques for analyzing and mining networks that have been suggested for a range of real-world networks~\cite{takes2014algorithms}. 
It turns out that the structure of the considered global corporate network has a power law degree distribution, meaning that the number of nodes with very few connections is large, whereas there are a smaller number of hub-like nodes with a very high degree. 
Moreover, nodes cluster together, forming a larger than random number of closed triangles of connections. 
Despite the low density of the network, the average distance (number of hops) between two nodes is relatively low, altogether referred to as the small world property~\cite{kleinberg2000small,kogut2012small}. 
This paper focuses on centrality measures, techniques commonly employed in small world networks for assessing the importance of a node with respect to the other nodes, based on the structure of the network. 
Well-known examples of such measures that originate from the field of social network analysis are degree centrality, eigenvector centrality, closeness centrality and betweenness centrality~\cite{Borgatti2006466,brandes2007centrality}. 

Networks of interlocking directorates have been studied for over 100 years, and there is an extensive body of literature discussing the causes and consequences of board interlocks, see for example the excellent overview in \cite{mizruchi1996interlocks}. 
In board interlock networks, node (firm) centrality is widely considered as an indication of a powerful or at least advantageous position~\cite{pfeffer1978external,stokman1985networks}. 
An extensive body of literature discusses the relationship between the economic performance of a firm and centrality~\cite{andres2013busy,croci2014economic,hillman2003boards,larcker2013boardroom,mariolis1982centrality}. 
However, this literature has found diverse outcomes in different countries when it comes to the precise relation between centrality and firm performance. 
We adopt the argument that the ordering of the nodes determined by a centrality measure is an indicator of the economic order of power. 
Given that powerful firms are typically larger players in the economy, a comparison of centrality with an economic performance indicator may give us some indication as to which centrality measure is most representative for finding powerful actors in the global corporate network. 
So, in the global corporate network, we say that centrality gives an indication of the importance of a firm within the global system of corporate control.

Although globalization has led to a world-wide connected network of firms, if a standard hierarchical community detection algorithm (to detect groups of nodes that are more connected with each other than with the rest of the network) is applied to the global corporate network, the resulting communities have a clear regional character~\cite{heemskerktakes2015,heemskerktakes2016soc}. 
In line with earlier studies~\cite{carroll2002there}, this reveals that the footprint of the national networks is still visible in the global network.
The fact that the global network is actually comprised of multiple smaller national networks indicates a so-called 
multi-level structure~\cite{kivela2014multilayer,lazega2015multilevel}. 
This has important implications for the use of centrality as an indicator of firm prominence.

First, firms have a certain central or less central position within the entire (global) network, but also within the partition of their respective national network. 
And these two may very well differ: a firm can be central in a national network but relatively peripheral in the global network.
A key methodological question addressed in this paper is therefore how centrality measures can be interpreted, compared and understood on the various global and national scales of the network using quantitative measures.
A comparison between local and global centrality rankings is far from trivial, firstly because the rankings that we compare are not of equal length, and second because one (the national) ranking's nodes are always included in the other (global) ranking. 
We will survey existing metrics for comparing centrality rankings and propose two new methods to compare such rankings in Section~\ref{sec:centralitycomparison}. 
The goal of these new metrics is to provide additional insight in how power and control at the national level of a country's corporate network is \emph{persistent} and \emph{dominant} at the global level. 

Second, as a result of the aforementioned multi-level structure we may be interested in how certain sets of firms, for instance those that are domiciled within a particular country, are embedded in the global network.
For this we will look at the differences between local and global centrality measures, as it provides insight in how the considered partition (national network) is embedded in the full network (globally). 
So, in addition to comparing descriptive static topological network properties between countries, as is also done for example in \cite{burris2012search}, we attempt to better understand the embeddedness of countries in the global corporate network based on the relation between local and global centrality. 

The three contributions of this paper are as follows. 
First and foremost, we discuss, compare, propose and evaluate existing as well as two new methods for quantitatively comparing centrality rankings at multiple scales of a network, such as the global and national scales in the considered corporate networks. 
Second, we conduct a cross-country comparison, comparing the topology of the global network to and between various national networks, including an analysis of how centrality relates to economic prominence indicators at the local and global scale. 
Third, using the newly proposed metrics and the insight in the local topologies, we are able to quantitatively compare, classify and rank countries based on their position within the network structure of the global economic order. 

The remainder of the paper is organized as follows. 
After discussing related work in Section~\ref{sec:relatedwork}, Section~\ref{sec:preliminaries} formally defines the considered corporate network at a local and a global scale. 
It furthermore gives a short review of the various centrality measures that we consider, before discussing and proposing various ways of comparing them at different levels of the network in Section~\ref{sec:centralitycomparison}. 
Next, the topology of our global corporate network dataset as well as the various national networks are investigated in Section~\ref{sec:dataset}. 
We apply centrality measures to our board interlock network in Section~\ref{sec:experiments}, specifically to better understand differences between centrality within the national networks and the global network. 
To do so, we experiment with the new comparison metrics proposed in the preceding section, using the obtained results to conduct an in-depth cross-country comparison. 
Finally, conclusions and suggestions for future work are given in Section~\ref{sec:conclusion}.


\section{Related Work}
\label{sec:relatedwork}

In this section we briefly survey literature on the analysis of corporate networks and board interlock networks, as well related work on centrality measures and means of comparing them. 
Finally, we discuss literature that deals with applying centrality specifically to board interlock networks. 

Apart from interlocking directorates, \emph{corporate networks} can model relationships between firms based on a number of different types of ties, including trade~\cite{wilhite2001bilateral}, borrowing and lending of money~\cite{battiston2016complexity} and ownership, creating a network in which two firms are linked if one firm owns a certain percentage of another firm~\cite{vitali2013community,vitali2011network}. 
In corporate networks, \emph{community detection} algorithms, to find groups of firms that are more connected with each other than with the rest of the network, are frequently applied~\cite{heemskerktakes2015,piccardi2010communities,vitali2013community}. 
For both board interlock as well as ownership networks, it has been suggested that the communities that arise from the global corporate network have a clear regional character. 

\emph{Interlocking directorates}, i.e., the fact that a person sits on two or more corporate boards, are of great interest to scholars from a variety of disciplines, including  political science, sociology, business administration, and more recently, network science. 
Together, interlocking boards connect the top decision making bodies of our economies in a ``social'' corporate network. 
These networks have been an object of study for over 100 years, dating back to the early 20th century, featuring the 1905 study by Jeidels~\cite{jeidels1905} of the board interlocks between German banks  and industrial firms. 
Corporate governance networks kept inspiring researchers throughout the 20th century. 
As described in \cite{mizruchi1996interlocks}, in an extensive body of literature the causes of interlocks were attributed to collusion, cooptation and monitoring (for example banks keeping an eye on firms they invested in), legitimacy (hiring board members with a particular reputation in a certain area that is of importance to the firm), individual career advancement and social cohesion (social ties among the upper class). 
It was established that these networks of board interlocks furthermore facilitate the spread of governance routines and practices, the exchange of resources, communication and the dissemination of new ideas~\cite{burris2005interlocking}. 
Beyond the boardroom, the role of other types of social ties between directors was also found to be of significant influence on a number of the aforementioned aspects~\cite{barnes2015structural}.
The consequences of well-connected boards are typically described in terms of corporate control and power, the embeddedness of firms within some economical system and to some extent, firm performance~\cite{larcker2013boardroom,mizruchi1996interlocks}. 
Studying board interlock networks furthermore gave rise to a debate on the existence of a transnational capitalist class~\cite{carroll2002there}, and differences in behavior between nationally and transnationally oriented boards were shown for example in \cite{murray2014evidence}. 

Centrality has long been a basic concept in the study networks of interlocking directorates, in the beginning focusing mainly on degree centrality.  
As social network analysis gained more popularity, new centrality measures were proposed and applied to understand networks of interlocking directorates~\cite{fennema1982international}, for example to see differences between banks and nonbanks~\cite{mariolis1982centrality}.
In \cite{hillman2003boards} it is argued that the function of monitoring and the provisioning of resources of well-connected board has an effect on firm performance. 
However, when it comes to the precise relationship of firm performance and topological board interlock network measurements, the results are diverse. 
Correlations between centrality and economic performance are frequently demonstrated, but differ in strength across studies. 
For example, in \cite{larcker2013boardroom} it was shown that higher node centrality in the United States results in better boardroom performance, measured using a number of economic performance indicators. 
In \cite{horton2012resources}, it was found that in the United Kingdom, the connectedness of directors and thus their boards is positively associated with firm performance, and a similar conclusion is drawn in \cite{cronin2004director} for director networks. 

There are also a number of works such as \cite{andres2013busy} that, using the case of Germany, suggest a negative correlation between firm performance and centrality. 
In \cite{croci2014economic}, using data on listed firms in Italy and a comparison with a number of previous works, it is argued that there are certainly significant differences between countries with respect to the correlation of board centrality and economic performance. 
Furthermore, the causal relationship between the connectedness of boards and the aforementioned consequences is not always clear, see for example the discussion in \cite{mizruchi1996interlocks}.  
In many papers it is left for future work to determine whether there is a causal effect, to study the differences between countries, or to scale up to sufficient data for a fair cross-country comparison. 
Such a comparison is difficult, because datasets of board interlock networks have different sources, and are frequently based on manually gathered data from annual reports. 
As a result, studies differ in terms of the number of firms that is studied and the point in time at which the study was done, making it hard to objectively compare results. 

In this paper we address a number of these issues, as we consider the largest $1$ million firms across the globe, allowing us to compare results with sufficient data in each country. 
The causal effects remain beyond the scope of this work, as we are foremost interested in understanding centrality at and between different national and global scales of the global corporate network network. 
Our work differs from studies such as \cite{everett1999centrality} in a sense that we still want to take the connectedness of the nodes within a particular partition of the full network into account, rather than merging all of the subset's nodes into one. 
To the best of our knowledge, this work is the first study in which the global corporate network is analyzed at such a large scale, particularly in the context of centrality, investigating the embeddedness of countries in the global network of corporate control. 


\section{Preliminaries}
\label{sec:preliminaries}

This section briefly describes the notation used throughout the paper to describe the various network aspects considered in our analyses.   
Apart from some general graph-theoretic terms and definitions, we formalize the two specific types of networks:  
the global corporate network consisting of firms across the globe, and national corporate networks of firms and interlocks from a particular country.
We furthermore give a definition of centrality measures in the context of corporate networks.

\subsection{Global corporate network} 
\label{sec:globalnetwork}

The \emph{global corporate network} is in fact a labeled undirected weighted network (graph) $G = (V, E, \omega, \phi)$. 
In this network, set of nodes $V$ nodes represent \emph{firms} (also referred to as boards, companies or corporations). 
The set of edges $E$ contains unordered pairs $\{u, v\}$ (with $u,v \in V$ and $u \neq v$) between two firms, denoting the fact that they share a common senior level director or officer, also referred to as a \emph{board interlock}. 
We use $n$ and $m$ for the number of nodes $|V|$ and the number of edges $|E|$, respectively. 
The value of $\mathit{deg}(v)$ stands for the degree, defined as the number of edges containing node $v$, so the number of interlocks of a particular firm's board. 

We aggregate possible parallel edges between firms (firms that have more than one common senior level director) by assigning a positive integer weight $\omega(e) \geq 1$ to each edge $e \in E$, indicating the number of common senior level directors between the two firms. 
A path is a sequence of nodes such that each subsequent pair in this sequence is directly connected via an edge in the network. 
A shortest path is a path of minimal length, and the length of such a path, starting at $u$ and ending at $v$, is a measure indicating the \emph{distance} between nodes $u$ and $v$, denoted $d(u, v)$. 
If for a particular maximal subset of nodes $V' \in V$ there exists a path between all pairs of nodes in $V'$, then $V'$ is called a connected component. 
The largest connected component is referred to as the \emph{giant component}. 
If two nodes are in different components, then there is no path between them and their pairwise distance is assumed to be infinite. 

The country $\phi(v)$ of a firm $v \in V$ is defined by the function $\phi \rightarrow H$ which maps each firm to one element of the set of countries in the world, denoted~$H$.

\subsection{National corporate network} 
\label{sec:nationalnetwork}

For each of the countries, a \emph{national corporate network} $G_h = (V_h, E_h, \omega_h)$ consisting only of the nodes and edges of a country $h \in H$ can be constructed. 
For the network of a certain country $h$, we define $V_h = \{v \in V : \phi(v) = h\}$ and $E_h = \{\ \{v,w\} \in E : \phi(v) = \phi(w) = h\ \}$. 
Note that $\cup_{h \in H} V_h = V$, but $\cup_{h \in H} E_h \subseteq E$, i.e., the union of the national node sets is the global node set, but the union of the national edge sets is not. 
This is due to the fact that transnational ties between firms of different countries are not included in any of the national networks, but are present in the global network. 

A different way of classifying countries within the global network would be to say that each country is actually a \emph{partition} of the network, where the edges within the partition are \emph{national ties} and edges between different partitions are \emph{transnational ties}. 
In theory, different unconnected components of the national network of some country may be indirectly connected through another country, so via transnational ties in the global corporate network. 
For all countries discussed in this paper, we found that the giant component of the national network was also part of the giant component of the full network. 

\subsection{Centrality measures}
\label{sec:centralitymeasures}

To determine the important actors in a network based on the structure of the network, centrality measures are commonly employed. 
A centrality measure $M$ assigns a function value $C_M(v)$ to each node $v \in V$, indicating the extent to which node $v$ has a central position in the network, based on the structure of the network. 
Here we ignore the edge weights. 
A centrality \emph{ranking} is simply a particular ordering on the set of nodes such that the for every subsequent pair $u, v$ in this ranking, $C_M(u) \geq C_M(v)$. 
The list of top-$k$ most central nodes of a particular network can thus be identified by sorting the set of nodes based on their centrality value, and then selecting the $k$ nodes with the highest centrality value. 
The four centrality measures considered in this work are:

\begin{description} 
	\item[Degree centrality] $$C_d(v) = \frac{\mathit{deg}(v)}{n -1}$$  
	\\
	\item[Closeness centrality] $$C_c(v) = \frac{1}{\sum_{w \in V}d(v,w)}$$
	\\
	\item[Betweenness centrality] $$C_{b}(u) = \!\! \!\! \mathop{\sum_{v,w \in V}}_{v \neq w, u \neq v, u \neq w} \!\!\!\! \frac{\sigma_u(v,w)}{\sigma(v,w)}$$
	Here, $\sigma(u,w)$ is the number of shortest paths from $u$ to $w$ and $\sigma_v(u,w)$ is the number of shortest paths that run through node $v$. 
	\\
	\item[Eigenvector centrality] $$C_e(v) = EV(v)$$
	This measure assumes that each node's centrality is based on the centrality values of the nodes that it is connected to, i.e., its neighbours. 
	It can be computed by iteratively setting $EV(v)$ of all nodes $v \in V$ to the average of that of its neighbours, where the initial values of $EV(v)$ are proportional to the degrees of the nodes, normalizing after each step.
\end{description}

\noindent
Although numerous other centrality measures have been suggested in literature, we believe that these four measures are the most common ones. 
More importantly, they each capture a different type of centrality. 
Respectively, they are based on a local property of the nodes (degree centrality), the average distance from the node to every other node (closeness centrality), the number of shortest paths that runs through a node (betweenness centrality) and the centrality of the node based on some iterative neighborhood-based propagation model (eigenvector centrality).
Each of the four measures can be normalized to the interval $[0, 1]$ by dividing it by the largest value over all nodes. 
This results in a situation in which a higher value indicates that the node is more central according to the considered measure. 
For a thorough review and analysis of the computational issues involved in determining the different centrality measures, we refer the reader to \cite{Borgatti2006466,brandes2007centrality}.

In this study, the use of centrality measures in both the national and the global networks is considered, resulting in two ``levels'' at which we can define centrality:

\begin{enumerate}
	\item \emph{Global centrality}: the centrality of a firm within the global corporate network, so considering all edges and ignoring the country attribute. 
	\item \emph{National centrality}: the centrality of a firm within the corporate network of one country, taking into account only the edges between nodes of the particular country. 
\end{enumerate}

\noindent
These two levels are essentially specialized cases of centrality in some original network and centrality computed just within a certain partition of that network. 
We are particularly interested in comparing the centrality values of firms on a global scale to centrality values on the national scale, so based on a certain country. 
The main question is then \emph{how} we should we actually compare national and global centrality rankings, which is the topic of the following section. 

\section{Centrality comparison techniques}
\label{sec:centralitycomparison}

Now that have the ``ingredients'' to model networks and to compute centrality at different scales of the corporate network, let us consider how we are going to interpret and compare them. 
The application of a centrality measure to a network dataset results in a ranking of nodes based on their structural position in the network. 
The advantage of using a ranking is that it counters the problems involved in bluntly comparing (averages of) centrality values between networks, as these values are inherently incomparable due to the different structure (size, density, clustering, etc.) of the compared networks. 
The main methodological question addressed in this section is therefore: \\

\noindent
\emph{How can we compare two different centrality rankings, possibly of different length, where the objects of one ranking are a subset of the other?} \\

\noindent
Ideally, we capture the relation between the two rankings in one properly normalized number, so that it is easy to compare results quantitatively. 
Indeed, given the sheer size of our network data, any manual comparison is infeasible, and the focus is therefore on an automated comparison approach. 
Although the issue of comparing centrality rankings is relevant for the analysis of any type of network dataset, we use the global corporate network as a running example. 
Figure~\ref{fig:runningexample} is a fictive ranking resulting from applying a particular centrality measure to the global network (left) and a second ranking that solely bases its centrality values on firms in Great Britain (right), indeed comparing global centrality and national centrality as discussed in Section~\ref{sec:centralitymeasures}. 
The remainder of this section considers three types of ranking comparison techniques: match-based measures, correlation-based measures and baseline-improvement  measures, introducing two new methods in the last two categories. 
The discussed measures are finally summarized in Table~\ref{tab:measures}. 

\begin{figure}[b]
\begin{multicols}{2}
\noindent \textbf{Global} \\
1. \texttt{US AT\&T INC.} \\
2. \texttt{US 7-ELEVEN INC.} \\
3. \texttt{GB ROYAL DUTCH SHELL} \\
4. \texttt{GB ERNST \& YOUNG EUROPE} \\
5. \texttt{KR SAMSUNG ELECTRONICS} \\
6. \texttt{GB PRICEWATERHOUSECOOPERS} \\
7. \texttt{CH RAIFFEISEN SCHWEIZ}  \\
8. \texttt{GB KPMG EUROPE}	\\
\noindent \textbf{Great Britain} \\
1. \texttt{GB ERNST \& YOUNG EUROPE} \\
2. \texttt{GB PRICEWATERHOUSECOOPERS} \\
3. \texttt{GB KPMG EUROPE} \\
4. \texttt{GB ROYAL DUTCH SHELL} \\
\hphantom{x} \\
\hphantom{x} \\	
\end{multicols}
\caption{Fictive example of top-$8$ most central nodes at two different scales.}
\label{fig:runningexample}
\end{figure}

\subsection{Matching-based measures}

One of the most trivial ways of comparing two rankings, both of length $n$, is to count the overlap in the top-$k$ (for some $0 < k < n$) of these two rankings, and expressing this count as a percentage of $k$. 
For example, in the ranking in Figure~\ref{fig:runningexample}, the overlap in the top-$4$ between the global and British ranking
is $2$, resulting in an overlap of $0.5$.
Although easy to compute and interpret and widely used for analyzing centrality in social networks, this way of comparing rankings does not take into account the order of the objects in the ranking: the matched firms (\texttt{GB ERNST \& YOUNG EUROPE} and \texttt{PRICEWATERHOUSECOOPERS}) are ranked in a different order in the two lists, which is not reflected in this simple measure of overlap between lists. 
Furthermore, when a ranking of a partition of the network is compared to a ranking of the full network (in this example the \texttt{GB} partition), the difference in the length of the lists is not considered, nor is attention given to the fact that firms may have the same rank because their centrality values are equal. 
Also, we note that for smaller values of $k$, the overlap is frequently equal to $0$ (in the example, this would be the case for $k=2$); the cut-off always is somewhat arbitrarily chosen, and may even cut the ranking right in the middle of a range of nodes with equal centrality values. 

\subsection{Correlation-based measures}
\label{sec:correlationmethods}

Computing the relation between two equal length rankings is traditionally done using the Spearman rank correlation coefficient, measuring the extent to which the relationship between two variables can be described using a monotonic function~\cite{spearman1904proof}. 
Especially when understanding centrality rankings in real-world network data, Spearman is frequently used~\cite{hahn2005comparative,yan2009applying}. 
Because we are interested in whether the rankings (and not so much the values) are equivalent, the measure of rank correlation is specifically suitable, as it is not subject to the size of the considered networks or the applied method of centrality value normalization. 
The advantage of applying Spearman rank correlation is that the exact difference in ranking between all pairs of nodes in both sets is taken into account. 
Note that if we are not interested in the difference in ranking but merely in whether or not the distinct pairs of nodes in the different lists are correctly ordered, we could also have used Kendall's tau, measuring the relation between the number of concordant (correctly ordered in both lists) and disconcordant pairs of nodes. 

Compared to matching-based methods, the advantage of correlation-based methods is that they can take the order of objects into account, and are not biased by some arbitrary cut-off. 
They are furthermore able to cope with equal centrality values. 
In such cases, so when objects in the data have identical values, Spearman typically assigns a rank to these objects that is set to the average of the rank range these objects are positioned in. 
This ensures that the sum of all ranks remains intact. 
The only problem is that traditionally, to compute a rank correlation, the rankings need to be of equal length. 

For comparing local and global centrality rankings, we propose a new metric called \mbox{\emph{\textbf{centrality~persistence}}}, and define it for some partition $S \subseteq V$ as the Spearman rank correlation $rc$ between the centrality values of the nodes in $S$ computed based on the network of only the nodes in $S$, compared to the centrality values of $S$, but then computed as part of the full network's node set $V$ (both according to the same centrality measure $M$). 
Formally: 

$$ rc(S, V) = \textrm{Spearman-correlation}(\overline{C_M^S}, \overline{C_M^V})$$

\noindent
Here, $\overline{C_M^X}$ is the vector of centrality values computed using centrality measure $M$ applied to the network consisting of all nodes (and edges between nodes) in set $X \subseteq V$, setting a null value for nodes not in $X$.
The function $\textrm{Spearman-correlation}(A, B)$ computes the Spearman correlation (as described above) for the overlapping (non-null in both) entries in $A$ and $B$. 

In our corporate board interlock networks, centrality persistence measures for a particular country the rank correlation between national centrality (within that country partition $S$) given by the vector $\overline{C_M^S}$ and global centrality given by the vector $\overline{C_M^V}$. 
So, it assesses the extent to which the ranking of firms in a country (partition) is maintained (persistent, value of $1$) or distorted (value of $-1$) in the full dataset.
In contrast to the the number of advantages of correlation-based methods described above, we note that the unequal size of the rankings is still a bit of a problem. 
Given that we only consider objects in the smaller list, we throw away information about the position of the objects in the full ranking. 
In the example in Figure~\ref{fig:runningexample} we would only be able to compare the relation between four British firms in the global ranking to the national ranking, regardless of where these British firms are positioned in the global ranking. 
Essentially, in comparing rankings based on correlation, we investigate if the order of some ranking in the partition is preserved in the full ranking, i.e., we measure its \emph{persistence}, but we do not yet know how central the nodes in this partition actually are in the full network. 

\subsection{Baseline-improvement measures} 
\label{sec:cprsection}

The matching-based and correlation-based techniques discussed above are not specifically designed to handle the comparison of rankings based on a partition and rankings based on the full network dataset.  
Here we propose an additional metric to solve this problem, by assuming that the considered partition is a random sample of the data, implying that the centrality values are simply the centrality values of nodes that are selected in a uniformly random way from the full set of nodes.
This in turn would mean that the nodes in the partition are assumed to be uniformly distributed over the ranking. 
Then we compute for the particular considered partition the extent to which the ranking of its nodes differs from a random distribution of these nodes over the full ranking. 
An alternative yet functionally equal definition would be to say that we are measuring the extent to which the subset is on average embedded in the middle, more near the top, or near the bottom of the ranking. 

Assume that a node $v \in S$ according to some centrality ranking has rank $r(v) \in [1,|V|]$ in the full rank of all nodes in $V$, where a rank of $1$ is highest (most central) and $|V|$ is lowest (least central). 
We then propose to compute the embeddedness of the partition in the full ranking using \emph{\textbf{centrality ranking dominance}} $rd(S, V)$, defined as:

$$ rd(S, V) = \frac{1}{2} - \frac{\sum_{v \in S} r(v)}{|S| \cdot (|V|+1)} $$

\noindent
If the value of $rd(S, V)$ is smaller than $0$, it means that the partition is on average less central (it has a higher than average ranking sum) in the full network, whereas a value higher than $0$ means it is on average more central (it has a lower than average ranking sum). 
A partition exactly in the middle of the ranking would have a value equal to $0$. 
More details as well as a proof of the validity of this metric for determining rank dominance is given in Appendix~\ref{appendixproof}. 

In the example in Figure~\ref{fig:runningexample}, the British firms have ranks 3, 4, 6 and 8, summing to 21, resulting in a value of $rd(V_{GB}, V) = \frac{1}{2} - 21 / (4 \cdot (8+1)) = -0.083$, meaning that the partition of British firms is less central than expected. 
On the contrary, the United States with firms at rank 1 and 2, summing to 3, has a centrality ranking dominance value of $rd(V_{US}, V) = \frac{1}{2} - 3 / (2 \cdot (8+1)) = +0.333$, indeed indicating that the US firms are ranked higher than expected.

\noindent
Centrality ranking dominance gives an indication of 
whether a partition has on average a lower or higher position in the full ranking, i.e., it indicates the \emph{dominance} of the partition. 
In Section~\ref{sec:experiments} we will use centrality ranking dominance together with the centrality persistence measure to compare the embeddedness of economic power orders within countries in the global network. 
Finally we note that together, these two measures cover the five important features of ranking comparison techniques surveyed above and summarized in Table~\ref{tab:measures}. 

\begin{table}[t]
	\caption{Features of different centrality ranking comparison techniques.}
	\label{tab:measures}
	\small
	\setlength{\tabcolsep}{4pt}
	\centering
	\begin{tabular}{lccc}
		\hline\noalign{\smallskip}
		& \textbf{Match-based} & \textbf{Correlation-based} & \textbf{Baseline-improvement} \\ 
		{Compare lists}	& $\times$ & $\times$ & $\times$ \\ 
		{Handle equal centrality values}	& & $\times$ & $\times$ \\
		{Order-preserving} 	&  & $\times$ & \\
		{Account for rank difference} 	&  & $\times$ & $\times$ \\
		{Nested inequal lengths rankings} 		& & & $\times$ \\	
		\noalign{\smallskip}\hline
	\end{tabular}
\end{table}


\section{Network topology}\label{sec:dataset}

In this section, we describe how our dataset was collected, discuss its quality and provide an overview of the (structural) properties of the resulting global corporate network, as well as topological characteristics of the largest $34$ national networks considered in our cross-country comparison. 

\subsection{Data collection}

The data used in this paper originates from the ORBIS database of Bureau van Dijk, which contains information on over 100 million public and private companies worldwide. 
An extraction of data on the largest firms worldwide that were registered as ``very large'' or ``large'', and as ``active'' was made in July 2013.
Only companies for which information was available on the country of domicile and the senior directors (board of directors, executive board, supervisory board or senior management) were selected.
We include all personal interlocks at both the senior management and board level, particularly because of the diversity in classifications of board and top management positions across the globe.
Because we are specifically interested in the network connecting corporate boards, we include only interlocks based on persons; firms that are registered as board members are disregarded. 
The result is a list of $971,891$ firms and a total of $3,272,523$ top executives. 

\subsection{Data quality}
\label{sec:dataquality}

The quality of the ORBIS data is more than reasonable, and for most larger firms, the error rate is low. 
Overall, ORBIS is recognized as a reliable data provider in a number of previous works~\cite{compston2013,vitali2011network}.
An extensive study of how representative this dataset is for the global economy based on, among other things, a comparison with the relative GDP, is given in \cite{heemskerktakes2015}. 
All in all, we are confident that our dataset captures the vast majority of significant worldwide economic activity. 

Given the large size of our dataset, a valid question is whether or not all interlocks are present at the same time, as the duration of the interlock is not explicitly included in the data. 
It should be noted that given that boards of larger companies meet every month and appoint their members for four years, on average re-electing a member once, there is close to one hundred opportunities for interaction between board members, facilitating the type of communication, interaction, exchange of ideas and resources known to come with board interlocks (see \cite{mizruchi1996interlocks} and the references in Section~\ref{sec:relatedwork}). 
In light of the low average pairwise distance between nodes (see later in this section), it is thus safe to assume that we can interpret the board interlocks as if they take place at the same time. 
It is anecdotally noted in literature that if boards meet every month, given the low average pairwise node-to-node distance of the small world interlock networks, an infectious disease affecting only directors could wipe out the majority of the corporate elite in well under a year~\cite{davis1991agents}. 

Finally, we note the presence of a number of administrative ties between firms in our dataset, for example as a result of how firms organize themselves through multiple legal entities or pyramidal structures of holding companies and corporate groups. 
Given the sheer size of the data there is no way to manually filter these ties given the available data. 
In line with previous work, we choose to leave these ties be, meaning that we have to take the presence of these ties into account when interpreting empirical results. 
For a lengthy discussion on the considerations around filtering these administrative ties, the reader is referred to \cite{heemskerktakes2015}.

\subsection{Network properties}
\label{sec:networkprop}

\begin{table}[!b]
	\caption{Global network properties.}
	\label{tab:dataset}
	\small
	\setlength{\tabcolsep}{4pt}
	\centering	
	\begin{tabular}{rl}
		\hline\noalign{\smallskip}
		\multicolumn{2}{c}{\textbf{Global corporate network}} \\
		\noalign{\smallskip}\hline\noalign{\smallskip}
			Nodes (firms)			& $391,967$ \\
			Edges (interlocks)		& $1,711,968$ \\
			Density					& $2.229 \cdot 10^{-5}$ \\
			Average degree	 		& $8.746$ \\
			Clustering coefficient 	& $0.755$ \\
			Connected components 	& $55,616$ \\ 
		\hline\noalign{\smallskip}
		\multicolumn{2}{c}{\textbf{Giant component}} \\
		\noalign{\smallskip}\hline\noalign{\smallskip}
			Nodes (firms)			& $238,859$ nodes ($60.9\%$)  \\
			Edges (interlocks)		& $1,533,030$ ($89.5\%$) \\
			Density					& $5.374 \cdot 10^{-5}$ \\ 
			Average degree	 		& $12.83$ \\		
			Clustering coefficient 	& $0.751$ \\
			Average distance		& $7.775$ \\
			Radius					& $18$ \\
			Diameter				& $34$ \\ 
	\noalign{\smallskip}\hline
	\end{tabular}
\end{table}

The list of firms and directors extracted from the source database is essentially a two-mode network. 
From this network, we can generate a projection of the firm-by-firm network by for each director adding an edge for every distinct pair of boards that this director sits on.  
This results in the global board interlock network that will be our main structure of interest. 
Table~\ref{tab:dataset} shows some statistics of this undirected network. 
Note that it only contains firms with at least one board interlock, thus filtering non-interlocking firms and reducing the number of firms (nodes) from $971,891$ to $391,967$.

\begin{figure}[t]
\begin{minipage}{.5\textwidth}
    \includegraphics[angle=0,width=\textwidth]{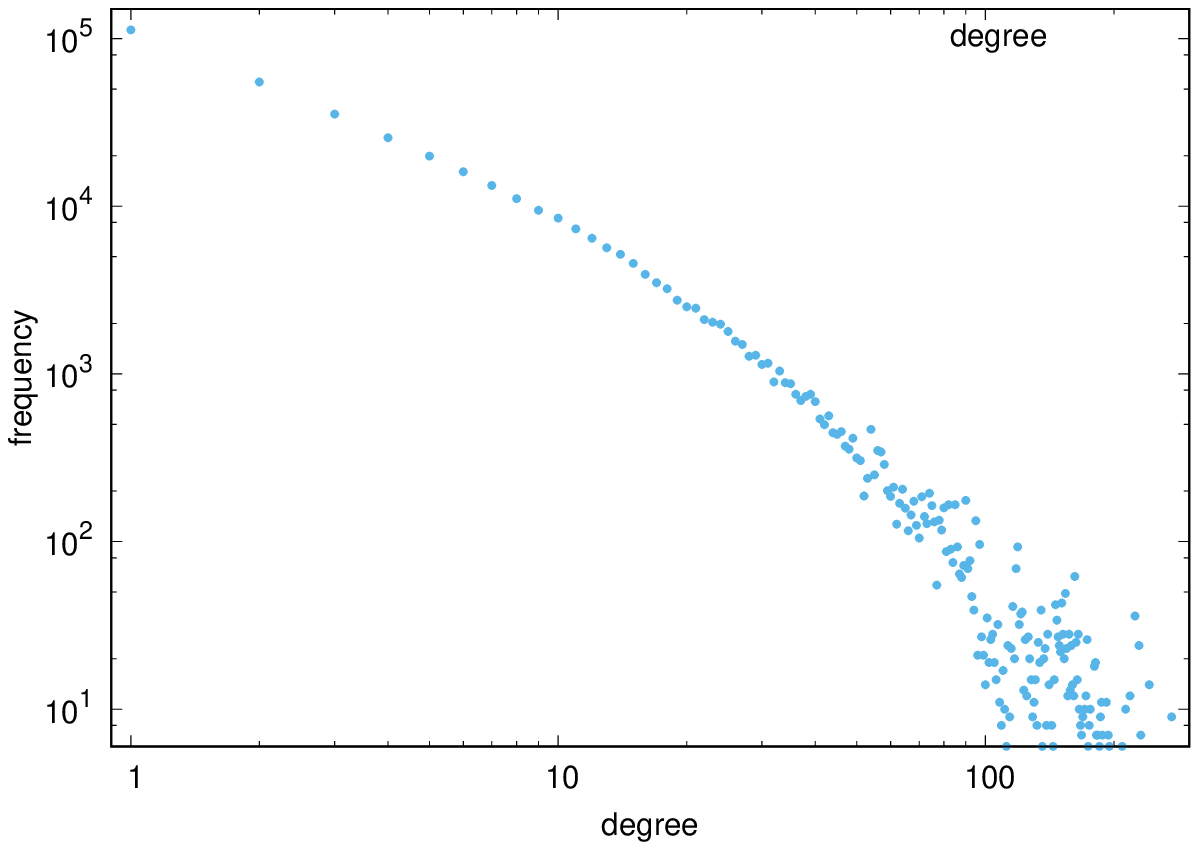} 
	\caption{Degree distribution.}
	\label{fig:degdist}
\end{minipage}
\begin{minipage}{.5\textwidth}
	\includegraphics[angle=0,width=\textwidth]{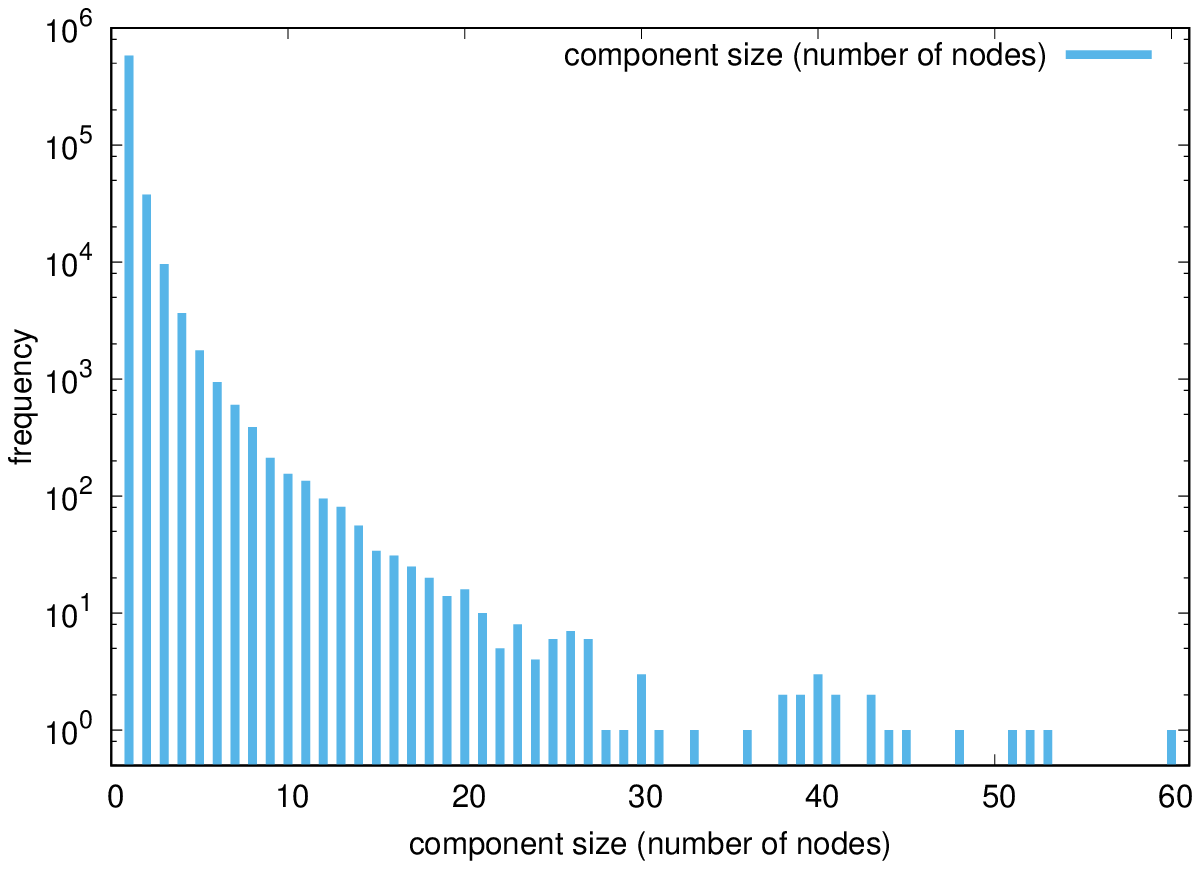}	 
	\caption{Component size distribution.}
	\label{fig:compdist}
\end{minipage}
\end{figure}

\noindent
The density, defined as the relation between the number of edges and the maximum number of edges, is low, as is common in many real-world networks. 
Figure~\ref{fig:degdist} shows the degree distribution of the network, which follows a power-law distribution and has a fat tail, again resembling many other real-world (social) networks~\cite{kleinberg2000small}. 
The fat tail is comprised of relatively few (note the logarithmic vertical axis) large firms with multiple economic entities that share a large number of senior directors. 
For example, some large accounting firms insist on all their partners being formal board members, demonstrating a type of administrative ties, as discussed in Section~\ref{sec:dataquality}. 
Although the effect of these administrative ties is small and merely local, it is something to take into account when studying a local metric such as degree centrality, as we will see later. 

\begin{figure}[b]
\begin{minipage}{.5\textwidth}
	\includegraphics[angle=0,width=\textwidth]{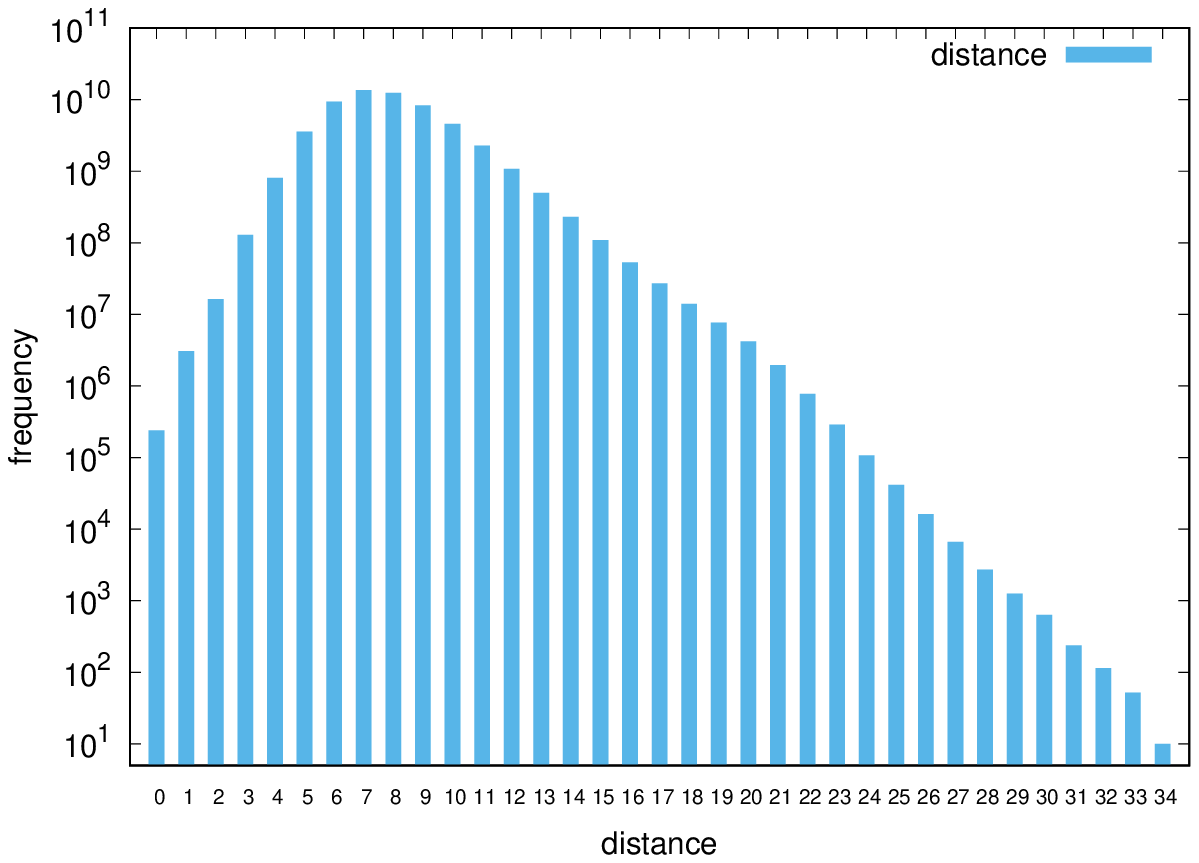} 
    \caption{Distance distribution.}   
	\label{fig:distdist}
\end{minipage}
\begin{minipage}{.5\textwidth}
	\includegraphics[angle=0,width=\textwidth]{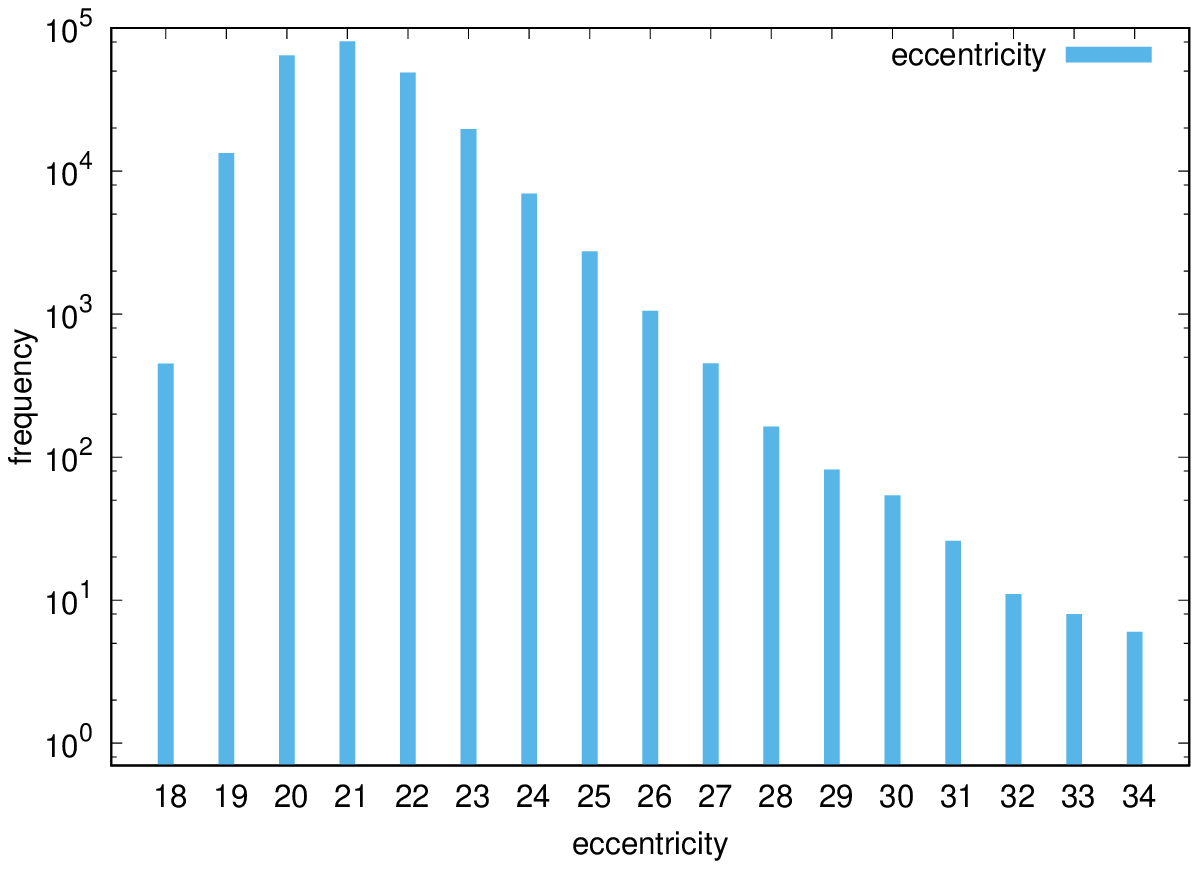}	
    \caption{Eccentricity distribution.}   
	\label{fig:eccdist}
\end{minipage}
\end{figure}

Next, the component size distribution for all $55,616$ components (excluding the largest component of size $238,859$) is given in Figure~\ref{fig:compdist}. 
We can observe that apart from the giant component, all other components are significantly smaller: they consist of at most $60$ nodes, with a clear peek at a size of $1$ to $5$.
The vast majority of these small components represent simple ``parent/subsidiary''-structures that do not share directors with the giant component, and are therefore hardly relevant for this study. 

Going from the full network to the giant component, the number of nodes drops with $39\%$ to $238,859$. 
However, the number of edges drops only by $10\%$, indicating that the majority of interlocking activity is captured in the $90\%$ of edges that reside within the giant component. 
Indeed, the other smaller components appear to be small isolated groups of firms mostly from the same country. 
The main focus of the remainder of this paper will therefore be on the giant component. 
We note that the high clustering coefficient of $0.751$ is partly attributed to the way in which the network was constructed: if an officer serves on more than one board, then all boards on which he serves are connected, automatically realizing a larger than normal number of closed triangles, which is exactly what is reflected by the clustering coefficient. 

Figure~\ref{fig:distdist} shows the distance distribution of the largest component (sampled over $100,000$ node pairs), with the average pairwise distance at $7.775$, which is consistent with other small world networks, but slightly higher than other social networks where the value is usually around $6$ or lower. 
The node eccentricity (length of the longest shortest path starting at a particular node) distribution over all nodes in the largest connected component is given in Figure~\ref{fig:eccdist}, starting at the radius ($18$) and ending at the diameter ($34$) and also has the familiar unimodal shape that is common in real-world networks~\cite{takes2013computing}.   

\begin{table}[!b]
	\small
	\caption{Network properties of giant components of the largest $34$ national networks.}
	\label{tab:countrydatasets}
	\small
	\setlength{\tabcolsep}{4pt}
	\centering	
	\begin{tabular}{clccccc}
		\hline\noalign{\smallskip} 
		\textbf{ISO2} & \textbf{Country}	 & \textbf{Nodes}	 & \textbf{Density}	 & \textbf{Clust. coeff.}	& \textbf{Avg. dist.} & \textbf{Transnat. factor} \\ 
		\noalign{\smallskip}\hline\noalign{\smallskip}
		AT	 & Austria			& 2,142		 & 0.00440	 & 0.273	 & 5.58	 & 0.79 \\ 
		AU	 & Australia		& 1,897		 & 0.00382	 & 0.085	 & 4.94	 & 0.58 \\ 
		BE	 & Belgium			& 3,264		 & 0.00254	 & 0.123	 & 5.17	 & 1.57 \\ 
		CA	 & Canada			& 5,439		 & 0.00146	 & 0.072	 & 5.20	 & 0.52 \\ 
		CH	 & Switzerland		& 999		 & 0.00620	 & 0.077	 & 4.78	 & 1.63 \\ 
		CN	 & China			& 891		 & 0.00475	 & 0.132	 & 5.80	 & 1.18 \\ 
		CO	 & Colombia			& 1,951		 & 0.00298	 & 0.090	 & 5.61	 & 0.34 \\ 
		DE	 & Germany			& 7,224	 	 & 0.00142	 & 0.320	 & 8.15	 & 0.63 \\ 
		DK	 & Denmark			& 4,517		 & 0.00229	 & 0.163	 & 5.61	 & 0.78 \\ 
		ES	 & Spain			& 11,102	 & 0.00143	 & 0.156	 & 6.30	 & 0.25 \\ 
		FI	 & Finland			& 2,626		 & 0.00294	 & 0.174	 & 5.52	 & 1.11 \\ 
		FR	 & France 			& 8,896	 	 & 0.00083	 & 0.170	 & 6.13	 & 0.77 \\ 
		GB	 & United Kingdom	& 32,962	 & 0.00067	 & 0.356	 & 6.63	 & 0.26 \\ 
		IE	 & Ireland			& 2,497		 & 0.01479	 & 0.178	 & 5.78	 & 0.39 \\ 
		IL	 & Israel			& 962		 & 0.01233	 & 0.065	 & 3.88	 & 0.21 \\ 
		IN	 & India			& 5,911		 & 0.00173	 & 0.047	 & 4.72	 & 0.20 \\ 
		IT	 & Italy			& 4,483		 & 0.00125	 & 0.198	 & 7.57	 & 0.88 \\ 
		JP	 & Japan			& 2,605		 & 0.00119	 & 0.113	 & 7.20	 & 0.21 \\ 
		KR	 & South Korea		& 2,802		 & 0.00174	 & 0.124	 & 5.83	 & 0.05 \\ 
		KY	 & Cayman Islands	& 642		 & 0.00693	 & 0.098	 & 5.40	 & 3.90 \\ 
		LU	 & Luxembourg		& 1,484		 & 0.00705	 & 0.196	 & 6.72	 & 1.55 \\ 
		MX	 & Mexico			& 931		 & 0.00852	 & 0.159	 & 4.31	 & 0.43 \\ 
		MY	 & Malaysia			& 7,878	 	 & 0.00398	 & 0.115	 & 4.50	 & 0.07 \\ 
		NL	 & Netherlands		& 6,083		 & 0.00271	 & 0.225	 & 7.61	 & 0.84 \\ 
		NO	 & Norway			& 8,963	 	 & 0.00130	 & 0.173	 & 5.69	 & 0.40 \\ 
		PT	 & Portugal			& 2,120		 & 0.00488	 & 0.138	 & 5.45	 & 0.56 \\ 
		RO	 & Romania			& 656		 & 0.00648	 & 0.189	 & 7.63	 & 1.92 \\ 
		RU	 & Russia			& 2,939		 & 0.00263	 & 0.102	 & 6.57	 & 0.08 \\ 
		SE	 & Sweden			& 6,656		 & 0.00166	 & 0.430	 & 6.40	 & 0.79 \\ 
		SG	 & Singapore		& 1,472		 & 0.00709	 & 0.080	 & 4.14	 & 0.90 \\ 
		TH	 & Thailand			& 981		 & 0.00555	 & 0.086	 & 4.90	 & 0.31 \\ 
		US	 & United States	& 24,802	 & 0.00024	 & 0.228	 & 6.71	 & 0.48 \\ 
		VN	 & Vietnam			& 1,393		 & 0.00558	 & 0.090	 & 4.44	 & 0.01 \\ 
		ZA	 & South Africa		& 963		 & 0.00837	 & 0.110	 & 4.10	 & 0.74 \\ 
	\noalign{\smallskip}\hline
	\end{tabular}
\end{table}

Table~\ref{tab:countrydatasets} lists the most important previously discussed measures and statistics for the $34$ considered national networks, identified by their ISO $2$-letter country codes. 
The differences in the number of firms per country are a result of the fact that some countries are larger and economically more developed than others. 
Properties that can be derived from other reported metrics have been left out for readability. 
The rightmost column titled ``Transnat. factor'' indicates the factor by which the number of edges increases if the transnational ties of this country are included. 
We observe in Table~\ref{tab:countrydatasets} that Belgium (BE), Switzerland (CH), China (CN), Finland (FI), the Cayman Islands (KY), Luxembourg (LU) and Romania (RO) stand out here with a relatively high number transnational ties. 
This may be a seen as evidence of the outward orientation of these countries. 
However, simply counting the number of transnational ties may be a too simple approach for determining this, as it merely considers the number of local transnational connections. 
Therefore we will try to better understand this observation using more complex metrics of embeddedness in Section~\ref{sec:experiments}. 

Most interesting to note about Table~\ref{tab:countrydatasets} is that in each of the countries, the average distance between nodes is low, typically much lower than the average distance of $7.775$ in the full global network. 
The average distance over all $34$ countries is $5.734$, and the weighted average distance (so, compensating for the number of firms in a country) is $6.204$. 
This may be a first hint to the fact that the national footprint of countries is still present in the global network: apparently the transnational ties in the full global network are not able to connect the national networks in such a way that the average distance remains as low as in the national networks. 


\section{Centrality experiments}
\label{sec:experiments}

In this section we perform experiments using the dataset described in Section~\ref{sec:dataset}.  First, in Section~\ref{sec:centralityexperiments} we directly use centrality measures in an attempt to characterize their stability by means of a comparison with firm prominence.
Then, we use the newly proposed metrics centrality persistence and centrality ranking dominance to understand the relation between national and global centrality in respectively Section~\ref{sec:comparenatglob} and Section~ \ref{sec:pcr}, ending with a number of general results and remarks in Section~\ref{sec:discussion}. 

\subsection{Comparing centrality measures}
\label{sec:centralityexperiments}

\begin{table}[b] 
	\caption{Correlation between centrality measures and with firm prominence (revenue), $n = 238,859$.} 
	\label{tab:centralitycorr}
	\small
	\setlength{\tabcolsep}{4pt}
	\centering	
	\begin{tabular}{lcccc}
		\hline\noalign{\smallskip}
		 & \textbf{Betweenness}	 & \textbf{Closeness}	 & \textbf{Degree}	 & \textbf{Eigenvector} \\ 
		\noalign{\smallskip}\hline\noalign{\smallskip}	
		\textbf{Betweenness}	 & 1.000	 & 0.430	 & 0.521	 & 0.356 \\ 
		\textbf{Closeness}	 & 0.430	 & 1.000	 & 0.495	 & 0.902 \\ 
		\textbf{Degree}	 & 0.521	 & 0.495	 & 1.000	 & 0.498 \\ 
		\textbf{Eigenvector}	 & 0.356	 & 0.902	 & 0.498	 & 1.000 \\ 
		\noalign{\smallskip}\hline\noalign{\smallskip}	
		\textbf{Firm prominence}	 & 0.192	& 0.109	& -0.046	& 0.064 \\
		\noalign{\smallskip}\hline
	\end{tabular}
\end{table}

After applying the four different centrality measures described in Section~\ref{sec:centralitymeasures} (betweenness centrality, closeness centrality, degree centrality and eigenvector centrality) to the giant component of the full global network, we can immediately observe that these measures are all correlated, in line with previous studies of centrality~\cite{Borgatti2006466,freeman1979centrality}. 
The first four rows of Table~\ref{tab:centralitycorr} show this correlation using the rank correlation coefficient for each pair of centrality measures. 
The results in Table~\ref{tab:centralitycorr} are in line with expectations with respect to how the measures are defined: betweenness centrality, measuring the extent to which a node has a brokerage-like position is least correlated with the other three measures. 
Eigenvector centrality and closeness centrality are highly correlated, indeed both taking the full network into account, considering each node contributing to the centrality value proportional to how far away it is. 
Each of the considered measures is correlated with degree centrality, being the most simple measure of centrality, merely counting the number of local connections. 
Given the low average pairwise distances in real-world networks, obviously the direct neighbourhood size (i.e., the degree) of a node has influence on any centrality measure.  

\begin{figure}[b]
	\centering
    \includegraphics[width=\textwidth]{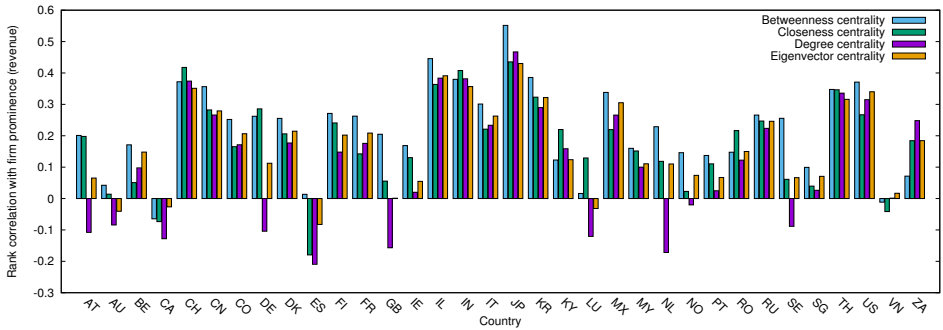} 
	\caption{Correlation between firm prominence (revenue) and national centrality for the $34$ countries in Table~\ref{tab:countrydatasets}.}
	\label{fig:countrycorr}
\end{figure}

Now that we have an idea of the relation between the measures themselves, let us see how the measures are related to firm prominence by comparing them with revenue. 
As argued in Section~\ref{sec:introduction}, this provides us with an indication of which measure is most representative for finding powerful actors in the global corporate network. 
The last row of Table~\ref{tab:centralitycorr} shows the results. 
We observe that of the four considered measures, betweenness centrality is most correlated with revenue at the global level, albeit a weak correlation. 
To further investigate, we can look at how firm centrality within individual countries correlates with revenue. 
For the largest $34$ countries (listed in Table~\ref{tab:countrydatasets}), the results of this experiment on a national level are shown in Figure~\ref{fig:countrycorr}. 
This figure essentially visualizes the same results as in the bottom row of Table~\ref{tab:centralitycorr}, but now for each of the countries, ignoring transnational ties.  
We see how in most countries, there is a weak to medium strong, but mostly \emph{positive} correlation between revenue and centrality, for almost each of the measures. 

The country-specific results are in line with results regarding firm performance that are presented for example for board interlock networks in the United States (US)~\cite{horton2012resources} and United Kingdom (GB)~\cite{larcker2013boardroom} and for director networks in \cite{cronin2004director}. 
At the same time they are conflicting with studies done on the board interlock networks of Germany and Italy presented in \cite{andres2013busy,croci2014economic}. 
A possible explanation for these differences may be found in the difference between data sources as well as the differences across studies in terms of which performance indicator the centrality metrics are compared to. 

Based on our experiments, we do establish that betweenness centrality seems the most suitable measure: apart from in Canada (CA) and Vietnam (VN) it is always weakly positively correlated with firm prominence. 
Indeed, we also saw this higher positive correlation in the full global network (Table~\ref{tab:centralitycorr}). 
Degree centrality and eigenvector centrality are in a few countries negatively correlated with firm prominence. 
Manually inspecting the data for some of the outliers based on degree centrality (the Netherlands (NL), United Kingdom (GB) and Canada (CA)) revealed multiple large densely connected communities of firms with very high degrees but no significant revenue, typical for the structures based on administrative ties discussed in Section~\ref{sec:dataquality}. 
These firms of course influence measures such as degree and eigenvector centrality. 
Noteworthy is the fact that again betweenness centrality in the Netherlands and United Kingdom shows no significant difference with other countries. 
In general, it is often said that degree centrality is a too simple measure of centrality, as it only measures connections locally, explaining its sometimes more erratic results when compared to revenue. 
As eigenvector centrality is ultimately biased towards high degree nodes, the various negative correlations for this measure are also understandable. 

There are vast differences between countries and how centrality is correlated with firm revenue, echoing the diverse results in previous work on the relation between firm performance and centrality. 
Apparently, at least for revenue, not in every country the well-performing firms that hold the most central positions in the board interlock network. 
Furthermore, we observe that across countries, the centrality values according to different measures can differ a lot. 
These findings suggests that, contrary to what is sometimes done~\cite{larcker2013boardroom}, assessing the centrality of a firm by averaging different centrality values, may at least in our case be to rigorous of an approach. 
Finally, we note that the correlation between revenue and centrality in the global network is much lower than in most of the national networks. 
In addition to our observations made at the end of Section~\ref{sec:networkprop}, this finding is a second piece of evidence suggesting that there are at least mechanisms at a local (national) scale that influence tie formation in the (global) board interlock network. 

\subsection{Centrality persistence}
\label{sec:comparenatglob}

\begin{figure}[b]	
	\centering
	\includegraphics[width=\textwidth]{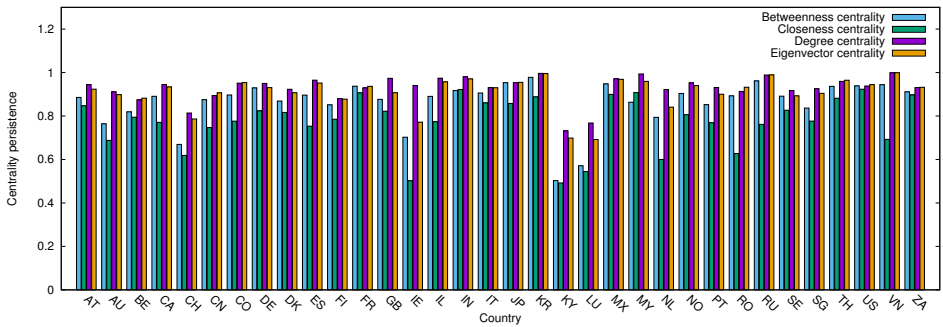}	
    \caption{Centrality persistence for the $34$ largest countries.}    
	\label{fig:countrynatglob}
\end{figure}

The second question that we aim to answer using our centrality experiments, is whether we can say something about how the order of firms on a national level is preserved in the global network. 
We proposed a measure for this in Section~\ref{sec:correlationmethods} called \emph{centrality persistence}, measuring the persistence of the order of a country's central firms within the global network. 
Recall that to compute this metric, we compare the rank correlation between firm centrality at the national level (so within a particular country) and the global level.
The results are shown in Figure~\ref{fig:countrynatglob} for each of the four considered centrality measures. 

Degree centrality persistence shows a noteworthy result. 
For most countries degree centrality values at a national and global scale are highly correlated: this is a direct result of the fact that degree centrality measures only direct connections, and nothing beyond that. 
If there is a big difference between national and global degree centrality, it means that there is a large number of transnational ties connected to firms in these countries, as this is the only local difference in edges between a country's nodes in the national and global network. 
Noteworthy are Switzerland (CH), Luxembourg (LU) and Cayman Islands (KY), as they  show the lowest degree centrality persistence values over all countries.   
These countries are frequently identified in literature as having large internationally oriented financial sectors~\cite{heemskerktakes2015}. 
The same observation holds to a lesser extent for Belgium (BE), China (CN) and Finland (FI). 
If we compare these results to the relative number of transnational ties listed in Table~\ref{tab:countrydatasets} in the column ``Transnat. factor'', we see a clear relation with degree centrality persistence. 
Indeed, degree centrality persistence measures indirectly the extent to which a country has a large number of transnational ties. 
The transnational factor (or a simple count of the number of such ties) is often used in literature to indicate globalizing countries based on networks of interlocking directorates, see for example \cite{burris2012search,veen2011national}, and degree centrality persistence essentially mirrors this aspect. 
In Section~\ref{sec:pcr} we will investigate more elaborate ways of determining whether each of these countries with a large number of transnational ties are actually dominant globalizing players within the world-wide economy. 	

As for the other measures, eigenvector and closeness centrality seem to again produce results that are very similar to degree centrality, as explained in the previous section. 
Let us now turn to the persistence of the ranking based on betweenness centrality, which we established as the relatively most stable indicator of prominence in the previous section.
Figure~\ref{fig:countrynatglob} shows that for Switzerland (CH), Luxembourg (LU) and the Cayman Islands (KY), analogously to degree centrality, we have low betweenness centrality persistence values. 
This suggests that the appearance of additional transnational ties had significant influence on the brokerage position of the firms in these countries as well. 
Ireland (IE) also stands out here with a relatively low betweenness centrality persistence value, yet it does not have a significantly larger number of transnational ties like the other three ``outliers''.

The general conclusion here is that most countries have a high centrality persistence value, meaning that overall, the order of firms is well-preserved in the full network. 
This in turn serves as a third piece of evidence that national footprints are still inherently present in the global network. 
However, what can we observe from the differences in centrality persistence between countries, i.e., when we compare the centrality persistence values between different countries?
Specifically, what happens when we compare the values to macro-economic indicators describing these countries? 


\begin{figure}[b]
	\centering
    \includegraphics[width=0.75\textwidth]{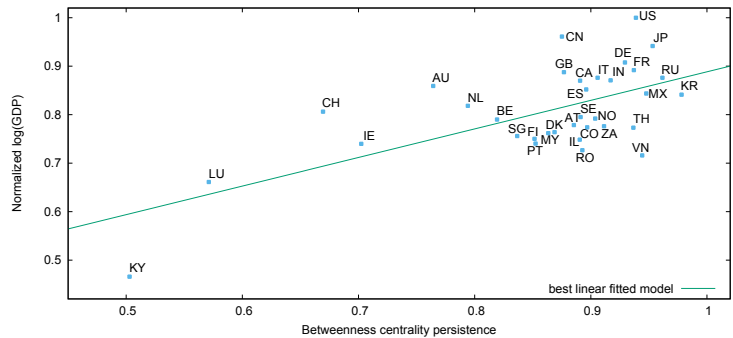}
	\caption{Betweenness centrality persistence vs. normalized log(GDP) for the $34$ countries in Table~\ref{tab:countrydatasets}.}
	\label{fig:persisgbp} 
\end{figure}

Most notably, we observe a correlation of $0.652$ (with a P-value of $2.91 \cdot 10^{-5}$) between betweenness centrality persistence and GDP. 
This means that to some extent, the higher the GDP (so, the larger the country's economy), the higher the centrality persistence value. 
This may indicate that, in general, countries with larger economies are also better at exhibiting the economic order of their firms within the global economy, i.e., they are better able to translate their power and control at the national level to the global level. 
The relation between GDP and betweenness centrality persistence is plotted separately for each of the $34$ countries in Figure~\ref{fig:persisgbp}.

We note that large economic powers such as the United States (US) and China (CN) are also exhibiting high centrality persistence, indicating that their national economic order of power hardly changes between the national and the global level. 
We may want to understand why certain countries with a similar GDP have roughly the same centrality persistence value. 
For example, in Figure~\ref{fig:persisgbp}, one may recognize two clusters of nodes; one with a group of firms strictly above the fitted line and one group below it. 
The first group consists mostly of developed countries such as Canada (CA), Germany (DE), France (FR), the United Kingdom (GB), Italy (IT) and Spain (ES), but also India (IN). 
The second group contains the larger Scandinavian countries, but nearby are also South Africa (ZA) and Colombia (CO). 
The countries within these groups obviously exhibit significant institutional differences. 
A natural question is how we can better understand the differences in centrality values of these seemingly similarly persistent countries. 
However, centrality persistence only gives a characterization of how prevalent a country's centrality ranking is, but it does not indicate whether the ranking is more persistent near the bottom or top of the global ranking, which we will investigate next. 

\subsection{Centrality ranking dominance}
\label{sec:pcr}

\begin{figure}[b]
	\centering
    \includegraphics[width=\textwidth]{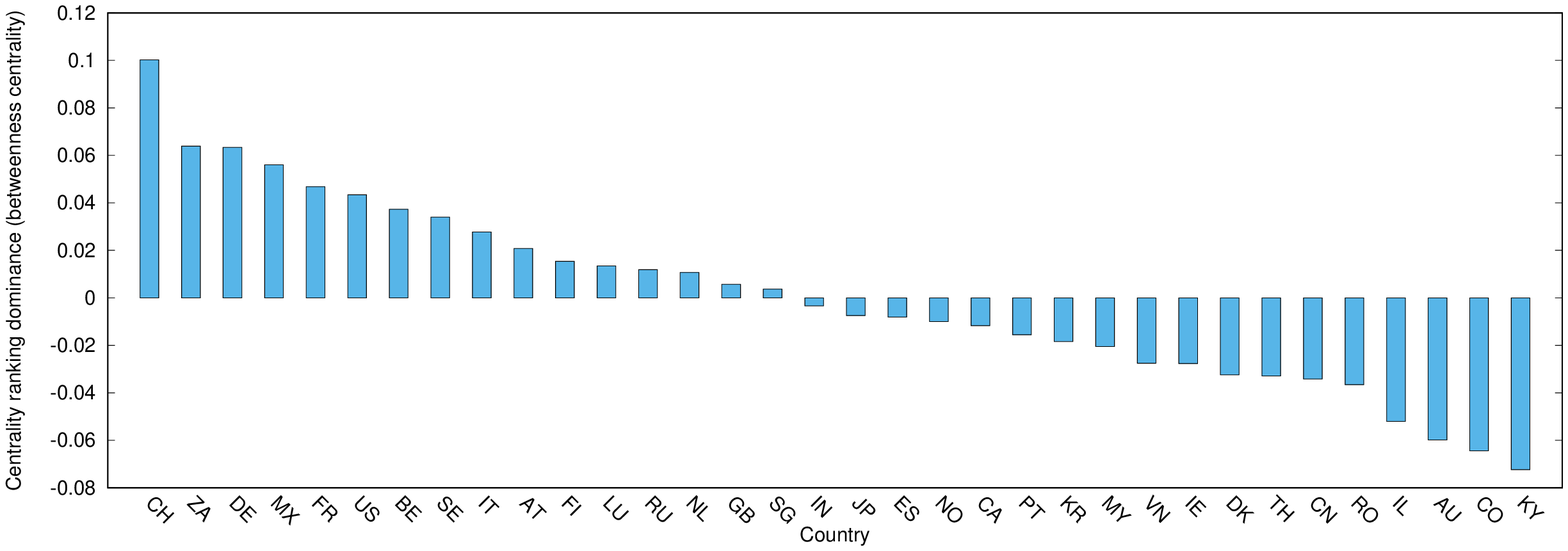}  
	\caption{Centrality ranking dominance for the $34$ countries in Table~\ref{tab:countrydatasets} (based on betweenness centrality).}
	\label{fig:countriesbi}
\end{figure}
\begin{figure}[b]
	\centering
    \includegraphics[width=0.75\textwidth]{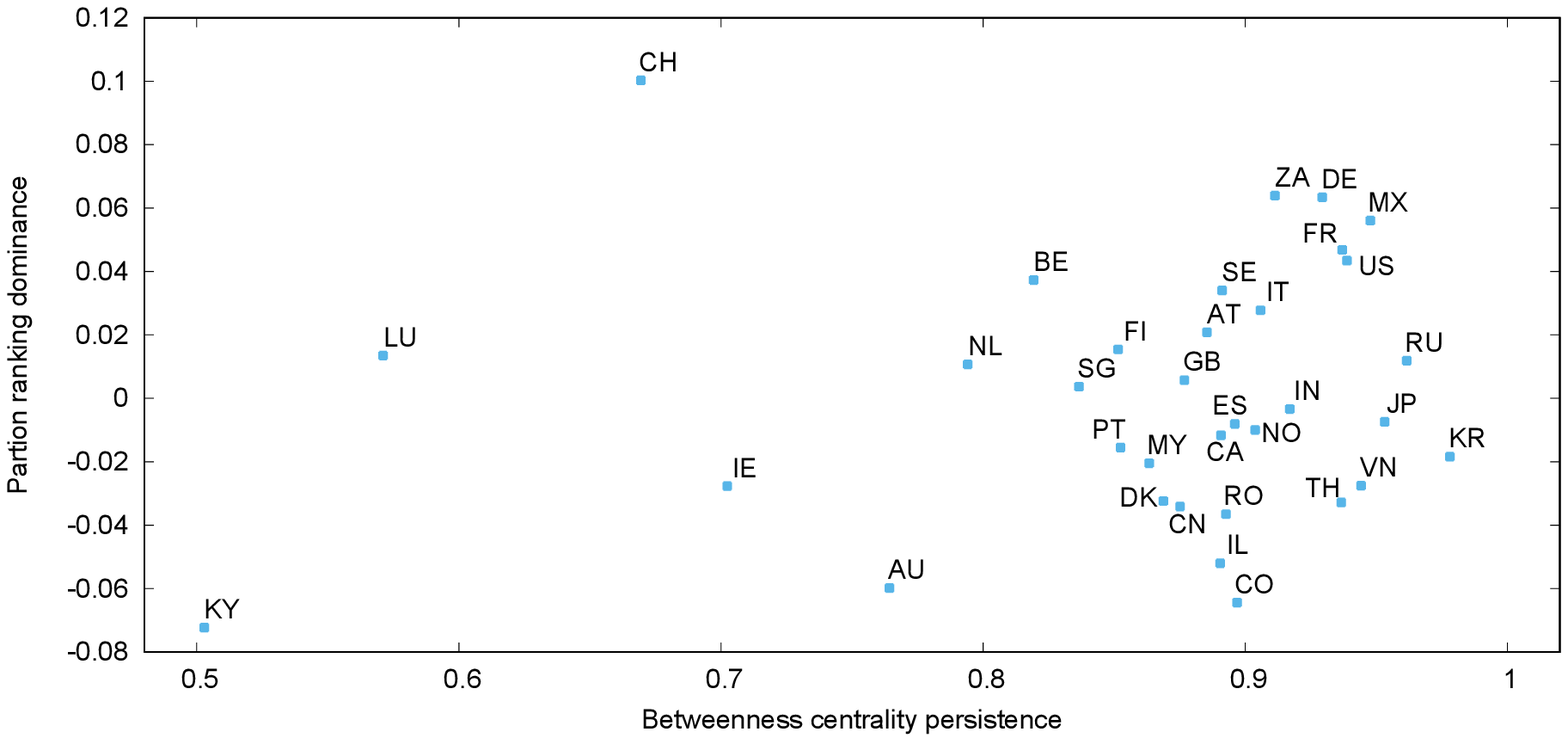}  
	\caption{Betweenness centrality persistence vs. ranking dominance.}
	\label{fig:persisdomi}
\end{figure}

We discussed in Section~\ref{sec:centralitycomparison} (and summarized in Table~\ref{tab:measures}) that the only downside of correlation-based comparison methods is that they do not measure very well whether a partition is more near the top or the bottom of a ranking of the nodes in the full network. 
Therefore, we now turn to experiments using the measure of \emph{centrality ranking dominance} suggested in Section~\ref{sec:cprsection}. 
The results for computing this measure for the $34$ largest countries are shown in Figure~\ref{fig:countriesbi}.  
Furthermore, a diagram showing both centrality ranking dominance and centrality persistence for each of the countries, is given in Figure~\ref{fig:persisdomi}. 

Recall that based on centrality persistence (see Figure~\ref{fig:persisgbp}) we were not able to distinguish so well between outliers such as Switzerland (CH) and the Cayman Islands (KY). 
Based on centrality dominance ranking we clearly can: these countries are quite the opposite in terms of the extent to which they dominate the centrality ranking. 
Although both are known for their large financial sectors, the Cayman Islands (KY) are frequently identified as transnationally oriented for administrative reasons or even fiscal benefits~\cite{heemskerktakes2015}, whereas Switzerland is known to be an influential actor in at least the European business community~\cite{heemskerkcommunity}. 
This is reflected in Figure~\ref{fig:countriesbi} with a low dominance value for the Cayman Islands and a high value for Switzerland. 

In the previous section, we also noted that Scandinavian countries such as Sweden (SE) and Colombia (CO) had a similar betweenness centrality persistence value. 
If we look at Figure~\ref{fig:countriesbi} and \ref{fig:persisdomi}, we see that they are quite different in terms of centrality ranking dominance: Sweden and Finland are more represented at the top of the ranking than Colombia. 
Typical well-developed western countries such as Switzerland~(CH), Germany~(DE), France~(FR), United States~(US) and Sweden~(SE) lead the ranking, but also South Africa (ZA) and Mexico (MX) are ranked higher. 
The latter two illustrate the brokerage position of these countries in the global economy.
Mexico connects Latin America and Europa (mostly through Spain), whereas South Africa connects the UK to Africa and parts of Asia. 
Along a similar line, with centrality ranking dominance we can now better distinguish the relatively lower ranked India (IN) from the western European countries that were observed to have a similar GDP and betweenness centrality persistence value in Section~\ref{sec:comparenatglob}.
Near the bottom of the ranking based on centrality dominance in Figure~\ref{fig:countriesbi} we also find China; apparently the Chinese firms are on average more located near the bottom of the global ranking. 
On the contrary, the United States (US) which based on centrality persistence appeared to mirror China, has a substantially higher rank and can thus be said to have a more dominant position in the economic order. 
This may be due to the fact that in Asia, board interlocks were mainly used to integrate firms operating in distinct business groups~\cite{fennema1982international,granovetter201019}.

Another example of added value of the centrality ranking dominance metric is the difference between China (CN) and Switzerland (CH). 
Based on the relative number of transnational ties in Table~\ref{tab:countrydatasets} we cannot easily distinguish between the two, but using the ranking dominance metric 
we can: Switzerland uses its transnational ties to obtain a dominant position in the global centrality ranking, whereas China clearly does not. 
Indeed, it is well-known that China is actively participating in the global economy, but is not yet as well integrated as other countries of the same size. 
The result for China is also to be expected: in previous work we already observed that especially China resides in its own subcommunity in the global board interlock network: China is the central country of the first persistent community discovered by a network community detection algorithm~\cite{heemskerktakes2015}. 
For Romania (RO), which also stood out in Table~\ref{tab:countrydatasets}, we observe a similar low ranking dominance value. 

The discussion and interpretation of results presented above demonstrates how in addition to centrality persistence and a count of the number of transnational ties, the measure of centrality ranking dominance allows us to better understand the relation between the national and global centrality rankings of different countries. 

\subsection{Discussion}
\label{sec:discussion}

The two discussed methods of centrality persistence and centrality ranking dominance allow centrality measures to be interpreted at different scales of the network. 
Together, they provide insight in the extent to which a partition is on the one hand able to preserve its centrality ranking in the full network (\emph{persistence}) and on the other hand whether the partition is on average more near the top or the bottom of the centrality ranking in the full network (\emph{dominance}). 
The results of these experiments result in a number of contributions to research on interlocking directorates.

Whereas in board interlock literature traditionally only the number of transnational ties (as shown in Table~\ref{tab:countrydatasets}) is used to characterize a country's participation in the global economy, we are now able to better distinguish between countries based on the extent to which they are holding a central positions at the national and global scale of the network. 
Furthermore, the two proposed measures can help understand the inward and outward orientation of a country and its firms.
Finally, they allow us to assess the embeddedness of countries based on the concepts of persistence and dominance, allowing more fine-grained insight in the extent to which a country is participating in the global economy. 

In studying the corporate board interlock network, we do not per se find evidence for key determinants based on intitutional patterns such as those demonstrated in \cite{veen2011national}, but instead we do see sector effects, in particular the financial sector and its influence on countries such as Switzerland, Luxembourg and the Cayman Islands. 
An extensive investigation of the precise cultural, geo-historical, political and economical differences and similarities of countries is nevertheless beyond the scope of this paper. 

\subsection{Final remarks}

We stress that the two proposed metrics have potential applications in many other types of networks that have attributes on the nodes that allow partitions of the full network to be studied. 
For example, in an online social network, the country of residence of a user could be used, whereas in scientific collaboration network the scientific field of the authors could be used to partition the network. 

We furthermore note that the experiments only report on significant correlations. 
For example, the reported correlations with revenue at a firm level and GDP at a country level are not near as strong for country-level demographics such as the population count or the GDP per capita of a country. 
Even more so, it should be noted that the proposed metrics operate independently of other topological metrics such as the number of firms (nodes), number of interlocks (edges), nor is there a noteworthy correlation between these basic network metrics and the proposed measures of persistence and dominance. 
All these findings support the general argument that a network approach truly provides additional insight compared to simply comparing countries using standard (macro-)economic attributes of these countries. 


\section{Conclusion}
\label{sec:conclusion}

We have investigated the concept of centrality in the global corporate board interlock network as well as within different national networks. 
Apart from the fact that centrality measures are obviously correlated with each other, we also find notable differences between countries. 
Most importantly, firm prominence and centrality do not always go hand in hand: large differences between countries have been demonstrated. 
In addition to previous work in which we showed how community detection revealed the footprints of national networks within the global network, this paper provides additional evidence for these regional effects based on both the network topology. 
First, we observe how on average, the $34$ largest national networks are more tightly connected (based on the average pairwise distance) than the global network. 
Second, we note that the relation between firm prominence and revenue is stronger in most of the national networks than in the global network, suggesting that there are mechanisms at the national level influencing the formation of board interlocks. 
Third, using the newly introduced measure of centrality persistence, the persistence of a national order of firms (ranked by centrality) within the global network is measured, showing high persistence values for a large number of countries. 
This means that for these countries, the economic order on a global scale is similar to that on the local scale, the third piece of evidence for the aforementioned national footprints. 
When comparing this persistence between countries, we find that there is a correlation with GDP: countries with larger economies are better at preserving their firm's central positions within the global corporate network than countries with smaller economies, although the persistence metric is not subject to the size (number of ties). 
The proposed centrality ranking dominance measures furthermore indicates the extent to which a country's firms are at a dominant position in a centrality ranking.  
Together, the two newly introduced measures of persistence and dominance give an indication of the extent to which a country's firms are at a central position at both a national and a global scal. 
The metrics allow us to get more fine-grained insight in the dominance and persistence of power and control of a country's firms at different scales of the global board interlock network. 

In future work, the applicability of our newly proposed measure of centrality persistence and ranking dominance could be further investigated. 
In corporate networks, non-geographical aspects, such as the sector in which the firm operates, could be used to define an alternative partitioning of the firms in the global network, allowing the influence of sectors to be investigated in a similar way as we did with countries.
This will however provide new challenges, as the sectors themselves are not necessarily as connected as countries. 
Furthermore, we plan to extend this research to the full corporate network consisting not only of large firms in the ORBIS database, but all firms, potentially increasing the number of nodes to over 200 million. 
One question is then whether or not considering all firms instead of just large firms influences the result, or that the majority of activity is already captured in our current selection of firms and directors. 
Finally, we plan to study the concept of centrality in corporate networks over time, as the global corporate network is constantly evolving through the formation of new board interlock ties. 


\subsection*{acknowledgements}
This research is part of the CORPNET project (see \url{http://corpnet.uva.nl}), which has received funding from the European Research Council (ERC) under the European Union’s Horizon 2020 research and innovation programme (grant agreement number 638946). \\
We thank Javier Garcia-Bernardo and Michiel Kosters for the fruitful discussions we had about this work, Fabio Daolio for feedback on an earlier version of this work, and finally the anonymous reviewers for their useful comments. 

\bibliographystyle{spmpsci}      

\bibliography{centralitysnam}  

\newpage


\appendix

\section{Centrality ranking dominance}
\label{appendixproof}

The measure of \emph{centrality ranking dominance} $rd(S, V)$ proposed in Section~\ref{sec:cprsection} computes the extent to which the ranking of nodes in a partition $S \subseteq V$ of the graph $G = (V, E)$ based on their centrality values is positioned near the top, middle or bottom of a ranking of the nodes, quantified using respectively a positive, zero or negative measure value in the range $[-\frac{1}{2}, \frac{1}{2}]$.  
In this short section we prove that this metric accomplishes exactly the functionality described above. 

Abstracting away from the network aspect, we have a set $V$ of objects that are ranked according to a centrality measure, meaning that we have a one-to-one mapping function $r$ that maps the nodes $v \in V$ to a rank integer $\in \{ 1, 2, \ldots, |V| \}$, where $1$ is the highest rank. 
We furthermore have a subset $S \subseteq V$ for which we want to determine whether the ranks $r(s)$ for all $s \in S$ are more dominant near the top, middle or bottom of the ranking.  
To do so, the measure of \emph{centrality ranking dominance} $rd(S, V)$ is formally defined as: 

$$ rd(S, V) = \frac{1}{2} - \frac{\sum_{v \in S} r(v)}{|S| \cdot (|V|+1)}$$

\noindent
If $S = V$, then 

$$\sum_{v \in S} r(v) = \sum_{v \in V} r(v) = \frac{1}{2} \cdot |V| \cdot (|V|+1),$$

\noindent
which means that 

$$ rd(S, V) = rd(V, V) = \frac{1}{2} - \frac{\frac{1}{2} \cdot |V| \cdot (|V|+1)}{|V| \cdot (|V|+1)} = \frac{1}{2} - \frac{1}{2} = 0,$$

\noindent
precisely indicating with a value of $0$ that $S$ is the middle of the ranking, i.e., $V$ is in the middle of $V$ itself. \\
If $|S| = 1$ with $S = \{s\}$, then

$$ rd(S, V) = rd(\{s\}, V) = \frac{1}{2} - \frac{r(s)}{|\{s\}| \cdot (|V|+1)} = \frac{1}{2} - \frac{r(s)}{|V|+1}.$$

\noindent
For $r(s) = 1$, $rd(\{s\}, V)$ approaches a value of $\frac{1}{2}$, and similarly for $r(s) = |V|$ the value approaches $- \frac{1}{2}$. 
The metric is in that sense symmetric. 
When $s$ is exactly in the middle of the ranking, for odd size $V$ we have $r(s) = \left \lceil{|V|/2}\right \rceil$, resulting in

$$rd(\{s\}, V) = \frac{1}{2} - \frac{\left \lceil{|V|/2}\right \rceil}{|V| + 1} = \frac{1}{2} - \frac{1}{2} = 0,$$ 

\noindent 
whereas for even size $V$ there would be no way to get a precise value of $0$ as one element can never be exactly in the middle of an even length ranking.
In such an even length ranking, $r(s) = |V|/2$ results in a value slightly lower than $0$ and $r(s) = (|V|/2) + 1$ gives a value slightly higher than $0$. 

\noindent
If the partition $S'$ contains more than one node, then for each node $s$ added to partition $S$ we compute its contribution to the metric in exactly the same way as outlined above for $|S| = 1$. 
This means that given the functionality for $|S| = 1$ above, we retain functionality for $|S| > 1$ as nodes $s$ are added:

$$
rd(S', V) 
= rd(S \cup \{s\}, V) 
= \frac{1}{2} - \left( \frac{\sum_{v \in S} r(v)}{|S| \cdot (|V|+1)} + \frac{r(s)}{|V|+1} \right) 
= \frac{1}{2} - \frac{\sum_{v \in S} r(v) + r(s)}{(|S|+1) \cdot (|V|+1)} 
$$

$$
= \frac{1}{2} - \frac{\sum_{v \in S \cup \{s\}} r(v)}{(|S|+|\{s\}|) \cdot (|V|+1)} 
= \frac{1}{2} - \frac{\sum_{v \in S'} r(v)}{|S'| \cdot (|V|+1)}
.$$

\hfill $\square$



\end{document}